\documentclass[aps,pra,twocolumn,superscriptaddress,showpacs]{revtex4}
\usepackage{amsbsy,latexsym}
\usepackage{amsfonts}
\usepackage{amssymb}
\usepackage[mathscr]{eucal}
\usepackage{graphics}

\newcommand{\emptysetz}{{\mbox{\boldmath$\emptyset$}}}
\newcommand{\emptysetsz}{{\mbox{\scriptsize\boldmath$\emptyset$}}}
\newcommand{\Lo}{{\mathscr L}}
\newcommand{\zero}{{\mathbf0}}
\newcommand{\one}{{\mathbf1}}
\newcommand{\az}{\mbox{\boldmath$a$}}
\newcommand{\bz}{\mbox{\boldmath$b$}}
\newcommand{\xz}{\mbox{\boldmath$x$}}
\newcommand{\xsz}{{\mbox{\scriptsize\boldmath$x$}}}

\newcommand{\id}{\openone}

\newcommand{\ket}[1]{\vert #1 \rangle}
\newcommand{\bra}[1]{\langle #1 \vert}

\begin{document}
\title{Multiple copy  2-state discrimination with individual measurements}

\author{A.~Ac\'\i n}
\affiliation{ICFO-Institut de Ci{\`e}ncies Fot{\`o}niques, Jordi
Girona 29, Edifici Nexus II, 08034 Barcelona, Spain}

\author{E.~Bagan}
\affiliation{Grup de F{\'\i}sica Te{\`o}rica, Universitat
Aut{\`o}noma de Barcelona, 08193 Bellaterra (Barcelona), Spain}

\author{M.~Baig}
\affiliation{Grup de F{\'\i}sica Te{\`o}rica, Universitat
Aut{\`o}noma de Barcelona, 08193 Bellaterra (Barcelona), Spain}

\author{Ll.~Masanes}
\affiliation{School of Mathematics, University of Bristol,
    University Walk, Bristol BS8 1TW, United Kingdom}
\affiliation{Dept. d'Estructura i Constituents de la Mat\`eria,
Univ. de Barcelona, 08028 Barcelona, Spain}

\author{R.~Mu{\~n}oz-Tapia}
\affiliation{Grup de F{\'\i}sica Te{\`o}rica, Universitat
Aut{\`o}noma de Barcelona, 08193 Bellaterra (Barcelona), Spain}

\begin{abstract}
We address the problem of non-orthogonal two-state discrimination
when multiple copies of the unknown state are available. We give
the optimal strategy  when only fixed individual measurements are
allowed and show that its error probability saturates the
collective (lower) bound asymptotically.
We also give the optimal strategy when adaptivity of individual
von Neumann measurements is allowed (which requires classical
communication), and show that the corresponding error probability
is exactly equal to the collective one for any number of copies.
We show that this strategy can be regarded as Bayesian updating.

\end{abstract}

\maketitle

\section{Introduction}

Measurement is a central tenet of quantum mechanics. As for any
sensible theory of nature,  it links abstract ideas to reality and
makes mathematical concepts truly physical.   In contrast to
classical measurements, which (ideally) have no demolishing effect
whatsoever, in the quantum realm any attempt to acquire
information from a system alters it to a degree proportional to
the gain of information. Moreover, this gain is
limited~\cite{Holevo}. Given a single copy of an unknown quantum
state it is usually impossible to determine it by performing any
conceivable measurement. Nevertheless, if an increasing number of
copies of such state is available,  our knowledge of the state can
also increase by the use of various measurement strategies, and
complete determination can be achieved in the asymptotic limit
when the number of copies goes to infinity.

Measurement strategies  involving multiple copies of a quantum
state fall into two categories:  collective and individual
(local), depending on whether a single measurement is  performed
on all the copies as a whole or the strategy consists of
individual measurements each of them performed separately  on a
single copy. Since the pioneering work of
Helstrom~\cite{helstrom}, and Peres and Wootters~\cite{pw}, it has
been repeatedly shown that {collective} strategies outperform
individual ones. This should not come as a surprise, since the
latter can be viewed as a subset of the former, which are
completely general and  unconstrained. Collective measurements,
however,  are difficult to implement experimentally, and a great
deal of effort go into designing optimal strategies involving only
individual measurements.  Common examples are: quantum
tomography~\cite{tomography}, and (local) adaptive
strategies~\cite{adaptative} (where the choice of each individual
measurement  is based on the outcomes of the previous), the two of
them in the context of quantum state estimation. The
state-of-the-art of these approaches can only compete with
collective strategies in the asymptotic limit.

Many practical applications, however, do not require a full
determination of a state. For instance,  to asses the security of
a key distribution protocol in quantum cryptography~\cite{cripto},
one gives full advantage to Eve, the eavesdropper. Hence, one
usually assumes she knows the set of possible states that will be
used in a secret transmission (e.g., in the B92
protocol~\cite{B92} this set consists of two non-orthogonal
states), and her task is to {\em discriminate}~\cite{chefles}
among them. She can follow two different approaches: use a
strategy based on  quantum hypothesis testing~\cite{helstrom}
(unconclusive discrimination), which gives the  lowest probability
of error, or  do unambiguous (or conclusive)
discrimination~\cite{unambiguous}, namely, adopt a strategy that
does not tolerate errors.

When the number of copies is greater than one (as is the case of a
non completely attenuated laser pulse which may be split in
several  identical single-photon states), the discussion above
concerning individual versus collective strategies becomes again
an issue. In this paper we focus our attention on this situation.
To be more concrete,  we will consider a hypothesis-testing
approach to (non-orthogonal) two-state discrimination under the
assumption that we have $N$ identical copies of the transmitted
quantum state. We will find the best adaptive strategy, i.e., a
particular case of strategies that use local operations and
classical communication (LOCC for short), and we will show that it
is optimal regardless the number of copies, in the sense that its
error probability and that of the optimal collective strategy are
{exactly} the same for {any} $N$. A similar  result was obtained
by Brody and Meister~\cite{brody} for Bayesian updating. Our
result could be seen as its extension to general adaptive
strategies. However, we will prove the remarkable result that the
whole class of adaptive strategies has actually a single element:
Bayesian updating.

If classical communication is not allowed, we show that optimality
holds asymptotically for the fixed measurement strategy named {\em
unanimity vote}, which we also present here.
\section{Preliminaries}
We will start by reviewing some known facts about two-state
discrimination, including a few technical details, which will help
us to introduce the notation.

\subsection{One copy}

By choosing the appropriate orthonormal basis, any two states
$\ket{\psi_0}$, $\ket{\psi_1}$ (which we will assume to be neither
orthogonal nor parallel) can always be written as
\begin{equation}
\ket{\psi_a}=\cos\theta\ket{x}+(-1)^a\sin\theta\ket{y} ;\quad a=0,1;
\label{psia}
\end{equation}
regardless the dimension of the Hilbert space $\mathscr H$ they
belong to, where the unit vectors $\ket{x}$, $\ket{y}$ are the
elements of the basis that span the plane $\mathscr P$ formed by
$\ket{\psi_0}$, $\ket{\psi_1}$. Now, we ask ourselves what  the
best measurement for discriminating between $\ket{\psi_0}$ and
$\ket{\psi_1}$ is. It can be defined in terms of two orthonormal
vectors, $\{\ket{\omega_1(0)},\ket{\omega_1(1)}\}$, which also
belong to ${\mathscr P}$, and thus  can be written as
\begin{equation}
\ket{\omega_1(a)}=\cos\left(\phi_0-a{\pi\over2}\right)
\ket{x}+\sin\left(\phi_0-a{\pi\over2}\right)\ket{y} .
\label{omega 1 copy}
\end{equation}

In our approach, by `best measurement'  we mean the measurement
that maximizes the probability of discrimination, $
P_1=\sum_{a=0}^1 q_a\, p(\az|a)=\sum_{a=0}^1\, p(\az,a)$ [or
equivalently the one that minimizes the error probability $\bar
P_1=1-P_1$). Here $q_a$ is the prior probability of $\ket{\psi_a}$
being (secretly) transmitted, $p(\zero |0)$ and $p(\one |1)$ are
the conditional probabilities of obtaining the outcomes $0$ or $1$
given that the unknown state is $\ket{\psi_0}$ or $\ket{\psi_1}$
respectively, and $p(\zero,0)$ and $p(\one,1)$ are the
corresponding joint probabilities. The subindex $1$ in
$\ket{\omega_1(a)}$ and in the probability of discrimination/error
emphasizes that so far we are dealing with just one copy of the
unknown state. Throughout this paper,  boldfaced random variables
will denote the outcomes of our measurement; thus, e.g.,
$p(1|\zero)$ is the (a posteriori) probability of the transmitted
state being $\ket{\psi_1}$ given that the outcome of  our
measurement is $0$. Using elementary quantum mechanics, the
conditional probabilities $p(\az|b)$ can be computed to be
$p(\az|b)=|\langle\omega_1(a)|\psi_b\rangle|^2=\cos^2[\phi_0-a\pi/2-(-1)^b\theta]$.
The optimal measurement and the corresponding probabilities of
discrimination and error are given by
\begin{eqnarray}
&\displaystyle \cos2\phi_0={q_0-q_1\over R_0}\cos2\theta ,\
\sin2\phi_0={q_0+q1\over R_0}\sin2\theta,&\label{cos2phi 1 copy}\\
&\displaystyle P_1={1\over2}(1+ R_0),\quad \bar P_1={1\over2}(1-
R_0),&\label{Pd 1 copy}
\end{eqnarray}
where $R_0 = [(q_0-q_1)^2+4q_0q_1\sin^22\theta]^{1/2}$. In terms
of the overlap between $\ket{\psi_0}$ and $\ket{\psi_1}$, defined
as $c\equiv|\langle\psi_0|\psi_1\rangle|=\cos2\theta$, the factor
$R_0$ can be written as
\begin{equation}\label{def R}
R_0=\sqrt{1-4q_0q_1 c^2}.
\end{equation}
In the simple case where $q_0=q_1=1/2$, we have $\phi=\pi/4$ and,
thus, $\ket{\omega(a)}=\{\ket{x}+(-1)^a\ket{y}\}/\sqrt2$, as one
would expect.

\subsection{Several copies. Collective measurements}

Let us next suppose that $N$ copies of either $\ket{\psi_0}$ or
$\ket{\psi_1}$ are available to us. In full analogy
with~(\ref{psia}) we define
\begin{equation}\label{Psia}
\ket{\Psi_a}=\ket{\psi_a}^{\otimes N}=
\cos\Theta\,\ket{X}+(-1)^a\sin\Theta\,\ket{Y} ,
\end{equation}
where $\ket{X}$, $\ket{Y}$ belong to a conveniently  chosen basis
of ${\mathscr H}^{\otimes N}$. In this situation
Eqs.~(\ref{cos2phi 1 copy}) and~(\ref{Pd 1 copy}) also hold if we
replace $\theta$ and $c$ with the corresponding  uppercased
variables $\Theta$ and $C=\cos2\Theta$.  In terms of the new basis
$\{\ket{X}, \ket{Y},\dots \}$, the vectors $\ket{\Omega(a)}$
($a=0, 1$),   which define the measurement on the $N$ copies in
full analogy with $\ket{\omega_1(a)}$,  are also given
by~(\ref{omega 1 copy}) (uppercasing $\omega$, $x$ and $y$). This
defines a {\em collective} measurement, since in general
$\ket{\Omega(a)}$ is not a product state. We obviously have
$C=|\langle\Psi_0|\Psi_1\rangle|=|\langle\psi_0|\psi_1\rangle^N|=c^{N}$
and thus conclude that the error probability  for this optimal
collective measurement is \cite{helstrom}
\begin{equation}\label{P col}
\bar P^{\rm col}_{N}={1-\sqrt{1-4q_0q_1c^{2N}}\over2} .
\end{equation}
Since $c<1$,  in the large $N$ limit we note that
\begin{equation}
\bar P^{\rm col}_{N}\simeq
{q_0q_1c^{2N}} .
\end{equation}

\section{Individual measurements}

\subsection{Fixed measurements}

If we are only allowed to perform the same individual measurement
on each of our $N$ copies, one could expect that the lowest
probability of error we can achieve is $\bar P^{\rm ind}_{N}\simeq
{\eta \,c^{N}}$, where the constant $\eta$ is not relevant for the
discussion here. This belief may stem from the widespread use of
the statistical overlap as a measure of distinguishability;  from
a statistical analysis of the problem at hand, one concludes that
the probability of error is bounded by $\lambda^N$, where
$\lambda$ depends on the specific individual measurement we are
performing. The statistical overlap is  a particularly convenient
choice of $\lambda$ (see below). Optimizing over all possible
measurements one finds that
$\lambda=c$ for two pure states.  
This bound is attained by a majority-vote strategy: we perform the
best individual measurement, given by~(\ref{omega 1 copy})
and~(\ref{cos2phi 1 copy}) on each copy and get $N_a$ times the
outcome~$a$. Once the measurement process is complete, we decide
in favor of the state $\ket{\psi_a}$ whose corresponding $N_a$ is
greatest.

However, there exist  tighter bounds for the exponential decrease
of the probability of error. The best one is known as Chernoff
bound~\cite{chernoff}, which for the problem at hand is given by
$\lambda=\min_\alpha\,\sum_b p(\bz|0)^\alpha p(\bz|1)^{1-\alpha}$,
where $0\le\alpha\le1$ (the statistical overlap is a particular
simplification of this expression obtained by setting
$\alpha=1/2$). We now note that if we assume $q_0>q_1$, with the
choice $\ket{\tilde\omega(0)}=\ket{\psi_0}$,
$\ket{\tilde\omega(1)}=\ket{\psi_0^\perp}$ for our measurement
we have $p(\zero|0)=1$, $p(\one|0)=0$, $p(\zero|1)= c^2$, and
$\alpha=0$ gives the absolute minimum (over all measurements and
over all values of $\alpha$) of the sum over~$b$ above. Thus,
$\bar P^{\rm ind}_{N}\simeq \eta c^{2N}$, as for collective
measurements.

There is a simple strategy that saturates the Chernoff bound:
unanimity vote. Let ourselves perform the measurement defined by
$\{\ket{\tilde\omega(a)}\}$ on each of our $N$ copies. If we
always obtain the outcome $0$  ($N_0=N$), we claim that the
unknown state is $\ket{\psi_0}$. However, if we obtain the outcome
$1$ once or more than once, we decide in favor of~$\ket{\psi_1}$.

The exact probability of error is straightforward to compute as
follows. Let us assume again $q_0>q_1$. If the unknown state were
$\ket{\psi_0}$, we would make no error. If the unknown state were
$\ket{\psi_1}$ (it happens with probability $q_1$), we would give
the wrong answer only if $N_0=N$, which happens with probability
$c^{2N}$. Hence, the probability of error would be ${q_1}c^{2N} $.
If $q_1>q_0$, we just exchange the subscripts 0 and 1 everywhere.
The error probability is then
\begin{equation}\label{PeindN}
\bar P^{\rm ind}_{N}=\min(q_0,{q_1})c^{2N} .
\end{equation}
We note that asymptotically $\bar P^{\rm ind}_{N}$ may be larger
than $ \bar P^{\rm col}_{N}$ only because of the prefactor
$\min(q_0,q_1)\ge q_0q_1$, which is not important in most
situations. This result has application in the assessment of the
security of some quantum cryptographic
protocols~\cite{preparation}.

\subsection{Adaptive measurements}\label{adaptive section}

So far, we have shown that the performance of individual and
collective  strategies is essentially the same for large ensembles
of identical states. We now show that if we are not restricted to
perform the same individual measurement on each copy, and we use
the information we are gathering to optimize these measurement
step by step, the overall performance is {\em exactly} the same as
for the optimal collective strategy, regardless the number of
copies of the unknown state. One could reach this conclusion by
using the algebraic results in~\cite{walgate} to trade
$\ket{\Omega(a)}$ for a set of product states similar to those
in~(\ref{prod stat}) below. We follow here a different approach
since we would like to present a constructive procedure within the
framework of probability.

We consider the simplest scenario where we perform always von
Neumann measurements on each individual copy. The final outcomes
are binary sequences or strings of length $N$, e.g.,
$\zero\one\one\cdots\zero\one$. Let us denote them by~$\xz$. The
strategy is designed in such a way that the last outcome (leftmost
binary digit in~$\xz$) determines whether our guess
is~$\ket{\psi_0}$ or~$\ket{\psi_1}$.  We have
\begin{equation}
P^{\rm ad}_{N}=\sum_{\xsz\in\Lo_{N-1}} \left\{q_0\,p(\zero\xz|0)+
q_1\,p(\one\xz|1)\right\} ,
\end{equation}
where 'ad' stands for adaptive, $\Lo_r$ is the set of binary
strings of length $r$, and~$\zero\xz$, $\one\xz$ are the strings
obtained by appending~$\zero$, respectively~$\one$, to the left of
the string~$\xz$.

Quantum mechanics tells us that the conditional probability of
obtaining the set of outcomes $\xz\in\Lo_r$ if the initial state
were $\ket{\psi_b}$ is
$p(\xz|b)=|\langle\Omega(\xz)|\psi_b^N\rangle|^2$, where
\begin{equation}\label{prod stat}
\ket{\Omega(\xz)}=\ket{\omega(\xz_r)}\otimes\ket{\omega(\xz_{r-1})}
\otimes\cdots\otimes\ket{\omega(\xz_1)},
\end{equation}
$\xz_k$ is the substring of length
$k$  ($0\le k\le r$) consisting of the $k$ rightmost digits of
$\xz$,
and
\begin{equation}
\ket{\omega(\az\xz)}=\cos\left(\phi_\xsz-a{\pi\over2}\right)\ket{x}
+\sin\left(\phi_\xsz-a{\pi\over2}\right)\ket{y} ,
\label{omega0}
\end{equation}
in analogy with~(\ref{omega 1 copy}).
Note that $\phi_\xsz$, the angle that defines the measurement $r+1$,
depends only on the list of outcomes, $\xz$, of the
previous $r$ individual measurements. One readily sees
that
$\sum_{\xsz\in\Lo_r}\ket{\Omega(\xz)}\bra{\Omega(\xz)}=\id
$
in ${\mathscr H}^{\otimes r}$,
which implies that
\begin{equation}
\label{q0q1}
\sum_{\xsz}p(\xz|b)=1,
\end{equation}
as it should be. We start with $r=0$ ($\Lo_0$  contains only the
empty string~$\emptysetz$) and set $\phi_\emptysetsz=\phi_0$, as
defined in Eq.~(\ref{cos2phi 1 copy}), which gives the optimal
measurement for one copy. For $r>0$, $\phi_\xsz$ will be
determined by requiring optimality step by step. We now can write
\begin{equation}\label{PdadN 2}
P^{\rm ad}_{N}=\sum_{a=0}^1\sum_{\xsz\in\Lo_{N-1}} p(\xz,a)
 |\langle\omega(\az\xz)|\psi_a \rangle|^2,
\end{equation}
where $p(\xz,a)$ is the joint probabilities of $\ket{\psi_a}$
being transmitted {\em and} we obtaining the (partial) outcome
list $\xz$. Namely, $ p(\xz,a)=q_a p(\xz|a)=q_a\;\prod_{s=1}^{r}
|\langle\omega(\xz_s)|\psi_a\rangle|^2 $ (assuming $\xz\in\Lo_r$).
Eq.~(\ref{PdadN 2}) can be written in terms of the angles $\theta$
and $\phi_\xsz$ using Eqs.~(\ref{psia}) and~(\ref{omega0}).
Maximizing over $\phi_\xsz$, we obtain
\begin{equation}
\cos 2\phi_\xsz={p(\xz,0)-p(\xz,1)\over
R(\xz)} c,
 \label{cos2phi'}
\end{equation}
where
\begin{equation}
\label{R}
 R(\xz)=\sqrt{[p(\xz,0)+p(\xz,1)]^2-
4p(\xz,0)p(\xz,1)c^2}  ,
\end{equation}
and we also have
\begin{equation}\label{sin2phi'}
\sin 2\phi_\xsz=\frac{p(\xz,0)+p(\xz,1)}{R(\xz)} \sin2\theta .
\end{equation}
Substituting
back in~(\ref{PdadN 2}) we obtain
\begin{equation}
P^{\rm ad}_{N}={1\over2}+{1\over2}\sum_{\xsz\in\Lo_{N-1}}R(\xz)  ,
\label{adaptive}
\end{equation}
where we have used that $\sum_{\xsz}p(\xz,a)=q_a$, which follows
from~(\ref{q0q1}). Eqs.~(\ref{cos2phi'}), (\ref{R})
and~(\ref{adaptive}) are analogous to Eqs.~(\ref{cos2phi 1 copy})
and~(\ref{Pd 1 copy}). Actually, the later can be seen as a
particular case
of the former if we define $p(\emptysetz,a)=q_a$ (this definition
is sensible, since the empty binary string means that no
measurement has yet been performed).

Having set up this framework, one can prove our main result. Namely,
that this adaptive strategy gives exactly
the same error probability as the optimal collective one for
any~$N$.
A straightforward calculation yields
\begin{eqnarray}
p(\az\xz,b)&=&{p(\xz,b)\over2}\left\{
1+(-1)^{a+b}\phantom{{c^2\over R(x_r)}}\right.\nonumber \\
&\times&\left. {p(\xz,b)+(1-2c^2)p(\xz,b\oplus1)\over R(\xz)}
\right\}, \label{recursion}
\end{eqnarray}
where $\oplus$ stands for sum mod~2,
and one can prove by induction the relation
\begin{equation}
q_0q_1c^{2r}[p(\xz,0)+p(\xz,1)]^2-
p(\xz,0)p(\xz,1)=0,
\label{assumption}
\end{equation}
for $\xz\in\Lo_r$,
which is obviously satisfied for $r=0$.

Using this relation in~(\ref{R}) and  recalling again
that~$\sum_{\xsz}p(\xz,a)=q_a$, we finally have the
result~$\bar P^{\rm ad}_{N}=\bar P^{\rm col}_{N}$.

It is
not difficult to show that
\begin{equation}
\cos 2\phi_\xsz=(-1)^{i_r} c\,
{\sqrt{1-4q_0q_1c^{2r}
\over1-4q_0q_1c^{2r+2}}} \ ,
\label{cos2phi simpl}
\end{equation}
where $i_r$ is the leftmost digit in $\xz\in\Lo_r$ and we have
used that ${\rm sign}[p(\xz,0)-p(\xz,1)]=(-1)^{i_r}$ [Note that
${\rm sign}(q_0-q_1)=(-1)^{i_0}$].

We immediately realize that the actual dependence of the
individual measurement~$r+1$ on previous outcomes is extremely
simple: it is just a function of the $r$-th outcome, i.e.,  of
$i_r$, rather than a function of the whole binary sequence~$\xz$.
In this sense, the optimal one step adaptive scheme is
`Markovian'. It is thus convenient to change the notation and
define $\phi_r\equiv\phi_{\xsz}$,
$\ket{\omega_{r+1}(a)}=\ket{\omega(\az\xz)}$, for $\xz\in\Lo_r$.
Eq.~(\ref{omega0}) becomes
\begin{equation}
\ket{\omega_{r+1}(a)}=\cos\left(\phi_r-a{\pi\over2}\right)
\ket{x}+\sin\left(\phi_r-a{\pi\over2}\right)\ket{y},
\end{equation}
where subscript $r+1$ refers to the measurement on copy $r+1$ and
$a=0,1$  is the corresponding outcome. Eq.~(\ref{omega 1 copy}) is
a particular case of this equation.

\subsection{Bayesian updating interpretation}

Finally, we would like to show that the adaptive strategy we have
presented has a natural interpretation as Bayesian updating (we refer
to~\cite{brody} for an alternative point of view). This, along with the results
of the previous section, proves that Bayesian updating is the {\em unique} solution
to the recursion relations~(\ref{recursion}) that define the best adaptive strategy.

Note that our knowledge of the system, which changes after each
measurement,  is encoded in the {\em a posteriori} probabilities
of $\ket{\psi_a}$ being the unknown state {\em given that} a
specific outcome has occurred when performing  the measurement on,
say,  the $r$-th copy. We will show below that these {\em a
posteriori} probabilities can be identified with $P^{\rm ad}_{r}$
and $\bar P^{\rm ad}_{r}$. Assuming this  for the time being, we
might be tempted to  take a Bayesian point of view and  use
$P^{\rm ad}_{r}$ to update our prior  probabilities for the next
measurement. Hereafter, we drop the superscript  `ad' to further
simplify the notation.

Suppose we have got the first copy of the unknown state. Our
optimal measurement will be defined by $\phi_0$ in
Eq.~(\ref{cos2phi 1 copy}). If we obtain the outcome $i_1=0$, we
will update our priors using the rule $q_0\to p(0|\zero)=P_1$, and
we will use again~(\ref{cos2phi 1 copy}) to optimize the
measurement on the second copy
 (similarly, if the first
outcome is~$i_1=1$, we will view $p(1|\one)=P_1$ as our prior
$q_1$ for the second measurement). Hence, the second measurement
is defined by $\cos2\phi_1=(-1)^{i_1}c\,
{|P_1-\bar{P}_1|(1-4P_1\bar{P}_1 c^2)^{-1/2}}$, and we obtain
that the discrimination (error) probability after the second
measurement is $P_2=[1+ (1-4P_1\bar{P}_1)^{1/2}]/2$
($\bar{P}_2=[1- (1-4P_1\bar{P}_1)^{1/2}]/2 $). This updating of
the prior probabilities can be carried out step by step until we
run out of copies. At step $r$ we will have
\begin{equation}
\cos2\phi_{r}=(-1)^{i_r}{|P_r-\bar{P}_r|\over R_r} c,
\label{cos2phirbayes}
\end{equation}
where by analogy with $R(\xz)$, we have defined $
R_r=(1-4P_r\bar{P}_r c^2)^{1/2} $, and we obtain
\begin{equation}
P_{r+1}={(1+ R_r)/2} .
\label{Pr+1}
\end{equation}
This leads to the recursion relation
\begin{equation}
R_{r+1}=\sqrt{1-(1-R_r^2)c^2},
\label{recursionbayes}
\end{equation}
whose solution can readily be seen to be $R_r=[1-4q_0q_1
c^{2r+2}]^{1/2}$, and we again find that $\bar P^{\rm
ad}_{N}=\bar P^{\rm col}_{N}$.

We still need to show that the {\em a posteriori} probabilities
indeed coincide with $P_r$. It suffices to prove it for the case
$r=1$, where this statement amounts to $P_1=p(0|\zero)=p(1|\one)$.
This result follows from the obvious formula
\begin{eqnarray}
&P_1=p(0|\zero) p(\zero) +p(1|\one) p(\one),&\\[-.85em]
&&\nonumber
\end{eqnarray}
where
$p(\bz)$ is the probability of obtaining the outcome $b$, {\em if}  the
`detailed  balance' relation
\begin{equation}
p(0|\zero)=p(1|\one)
\label{statement1}
\end{equation}
holds
for the optimal scheme. Let us prove this is the case.

Using Bayes formula we can cast~(\ref{statement1}) as
\begin{eqnarray}
&&{|\langle\omega_1(0)|\psi_0\rangle|^2 q_0\over p(\zero)}=
{|\langle\omega_1(1)|\psi_1\rangle|^2 q_1\over p(\one)}.
\label{statement2}\\[-.7em]
&& \nonumber
\end{eqnarray}
We further note that the probabilities of obtaining the outcome
$a$ can simply be written as:
$p(\az)=\sum_b|\langle\omega_1(a)|\psi_b\rangle|^2 q_b$.
Therefore, Eqs.~(\ref{statement1}) and~(\ref{statement2}) are
equivalent to
\begin{equation}
{|\langle\omega_1(0)|\psi_1\rangle|^2 q_1\over
|\langle\omega_1(0)|\psi_0\rangle|^2 q_0}=
{|\langle\omega_1(1)|\psi_0\rangle|^2 q_0\over
|\langle\omega_1(1)|\psi_1\rangle|^2 q_1}.
\end{equation}
This, in terms, is equivalent to
\begin{equation}
(q_0-q_1)\sin2\phi\;\cos2\theta=(q_0+q_1)\cos2\phi\;\sin2\theta ,
\end{equation}
which obviously holds for the optimal strategy [see
Eq.~({\ref{cos2phi 1 copy}})], and concludes the proof.

\section{Concluding remarks}

In summary.  Multiple-copy two-state discrimination strategies
based on individual measurements can be as good as the best
collective ones. For fixed measurements, this statement holds only
asymptotically. By relaxing this constrain and allowing Bayesian
updating, which is arguably the simplest, easiest to implement,
adaptive strategy,  the statement holds for {\em any} finite
number of copies.
Furthermore, our approach provides very simple recursion relations
[e.g., ~(\ref{cos2phirbayes}), (\ref{Pr+1}),
and~(\ref{recursionbayes})]  or even closed-form expressions
[e.g.,~(\ref{cos2phi simpl}); recall the change of notation
$\phi_r=\phi_\xsz$] for the angles $\phi_r$ defining the optimal
von-Neumann measurements and the discrimination/error
probabilities.

Finally, we would like to point out that the general adaptive set
up of Sec.~\ref{adaptive section}, where measurements are allowed
to depend on histories or lists of outcomes (rather than just the
very last outcome) has a unique solution which can be regarded as
Bayesian updating. Despite all this generality, the optimal
solution is as simple as can be.

\section{Acknowledgments}

We acknowledge financial support from Spanish Ministry of Science
and Technology project BFM2002-02588, ``Ram\'on y Cajal" grant,
2002FI-00373 UB grant, CIRIT project SGR-00185, and QUPRODIS
working group EEC contract IST-2001-38877.


\begin{thebibliography}{99}

\bibitem{Holevo} A.S.~Holevo, \textit{Probabilistic and Statiscal Aspects
                of Quantum Theory} (North Holland, Amsterdam, 1982).
\bibitem{helstrom} C.W. Helstrom,
\textit{Quantum Detection and Estimation Theory}.
 (Academic Press, New York, 1976).
 \bibitem{pw}A.~Peres and W.K.~Wootters, Phys.~Rev.~Lett.~\textbf{66}, 1119 (1991).
\bibitem{tomography}
         A.~G.~White, \textit{et al.}, Phys.~Rev.~Lett.~\textbf{83}, 3102 (1999);
          D.~F.~V.~James, \textit{et al.}, Phys.~Rev.~A~\textbf{64}, 052312 (2001);
          R.~T.~Thew, \textit{et al.}, Phys. ~Rev.~A~\textbf{66}, 012303 (2002);
          J.~B~Alepeter, \textit{et al.}, Phys.~Rev.~Lett.~\textbf{90}, 193601 (2003).
\bibitem{adaptative} D.~G.~Fisher, S.~H.~Kienle and M.~Freyberger,
        Phys.~Rev.~A \textbf{61} 032306 (2000);
        R.D.~Gill and S.~Massar, Phys.~Rev.~A \textbf{61}, 042312
        (2000);
        Th.~Hannemann {\it et al.}, Phys.~Rev.~A~\textbf{65}, 050303
        (2002); E.~Bagan, M.~Baig and R.~Munoz-Tapia,
             Phys.~Rev.~Lett.~\textbf{89}, 277904 (2002).
\bibitem{cripto} N. Gisin, G. Ribordy, W. Tittel and H. Zbinden,
Rev. Mod. Phys {\bf 74}, 145 (2002).
\bibitem{B92} C.~H.~Bennett, Phys.~Rev.~Lett. {\bf68}, 3121 (1992).
\bibitem{chefles} A.~Chefles, Contemp.~Phys.~\textbf{41}, 401 (2000).
\bibitem{unambiguous} I.D.~Ivanovic, Phys.~Lett.~A \textbf{123}, 257 (1987).
\bibitem{brody} D.~Brody and B.~Meister. Phys.~Rev.~Lett.~\textbf{76}, 1 (1996).
\bibitem{chernoff} T.M.~Cover and J.A.~Thomas, \textit{Elements of
Information Theory} (Wiley Series in Telecommunications, New York,
John Wiley \& Sons, 1991).
\bibitem{preparation} A.~Ac\'\i n et al., in preparation.
\bibitem{walgate} J.~Walgate  \textit{et al.}, Phys.~Rev.~Lett.~\textbf{85}, 4972
(2000);
S.~Virmani, {\em et al.}, Phys. Lett.~A~{\bf228}, 62 (2001).


\end{thebibliography}
\end{document}